\documentclass[12pt]{article}
\usepackage{latexsym}
\usepackage{cite}

\begin{document}

\begin{center}
{\Large \bf Hamiltonian structure of the complex Monge-Amp\`ere
equation} \\[4mm]
{\large \bf Y Nutku$^1$ and M B Sheftel$^2$}
\\[3mm] $^1$ Feza G\"{u}rsey Institute, PO Box 6, \c{C}engelk\"oy, 81220
Istanbul, Turkey \\
 $^2$ Department of Physics, Bo\u{g}azi\c{c}i University,
  34342 Bebek, Istanbul, Turkey
 \vspace{1mm}
\\ E-mail: nutku@gursey.gov.tr,
  mikhail.sheftel@boun.edu.tr
  \end{center}

\begin{abstract}\noindent
 We discover Hamiltonian structure of the complex Monge-Amp\`ere
equation when written in a first order two-component form. We
present Lagrangian and Hamiltonian functions, a symplectic form
and the Hamiltonian operator that determines the Poisson bracket.
\end{abstract}
AMS classification scheme numbers: 35Q75, 83C15

\section{Introduction}

In earlier papers \cite{nns,nsy} we presented complex
multi-Hamiltonian structure of Pleba\~nski's second heavenly
equation \cite{pleb}, which by Magri's theorem \cite{magri} proves
that it is a completely integrable system in four complex
dimensions. A first important step for obtaining this result was
the discovery of a Hamiltonian structure of the second heavenly
equation that was set into a two-component evolutionary form. In
this paper our goal was to discover a Hamiltonian structure of the
the complex Monge-Amp\`ere equation ($CMA$) as a first step to
obtaining its multi-Hamiltonian representation and hence its
complete integrability in the sense of Magri.

This paper is based on our unpublished results obtained two years
ago where we overlooked that, in fact, we had then obtained a
symplectic, and hence Hamiltonian, structure of $CMA$.

In section \ref{sec-1storder} we set the complex Monge-Amp\`ere
equation in a two-compo-\ nent first-order evolutionary form and
find the form of its Lagrangian that is appropriate for
Hamiltonian formulation. In section \ref{sec-Hamilton} we discover
a symplectic structure and Hamiltonian structure of this $CMA$
system.

\section{Complex Monge-Amp\`ere equation in first-order evolutionary form and its Lagrangian}
\setcounter{equation}{0}
 \label{sec-1storder}

The complex Monge-Amp\`ere equation has the form
\begin{equation}
u_{1\bar 1}u_{2\bar 2} - u_{1\bar 2}u_{2\bar 1} = \varepsilon
 \label{cma}
\end{equation}
where $u$ is a real-valued function of the two complex variables
$z^1 ,z^2$ and their conjugates $\bar z^1 ,\bar z^2$, the
subscripts denote partial derivatives with respect to these
variables. Here $\varepsilon$ is an arbitrary constant which means
in essence that either $\varepsilon = \pm 1$ or $\varepsilon = 0$,
though we shall not be interested here in the latter simple
special case.

$CMA$ (\ref{cma}) is a second-order partial differential equation,
so in order to discuss its Hamiltonian structure, we shall single
out a real independent variable, $t=2\Re{z^1}$, in (\ref{cma}) to
play the role of ``time", introduce its companion real variable
$x=2\Im{z^1}$ and change the notation for the second complex
variable $z^2=w$. Then (\ref{cma}) becomes
\begin{equation}
(u_{tt}+u_{xx})u_{w\bar w} - u_{tw}u_{t\bar w} - u_{xw}u_{x\bar w}
+ i(u_{tw}u_{x\bar w} - u_{xw}u_{t\bar w}) = \varepsilon .
 \label{cmareal}
\end{equation}
Now we can express (\ref{cmareal}) as a pair of first-order
nonlinear evolution equations by introducing an auxiliary unknown
$v = u_t$
\begin{equation}
\left\{
\begin{array}{l}
u_t = v \\
v_t = -u_{xx} + \frac{\textstyle 1}{\textstyle u_{w\bar
w}}\Bigl(v_wv_{\bar w} + u_{xw}u_{x\bar w} + i(v_{\bar
w}u_{xw}-v_wu_{x\bar w}) + \varepsilon\Bigr),
\end{array}
\right.
 \label{uv}
\end{equation}
so that finally (\ref{cma}) adopts a two-component form. For the
sake of brevity we shall henceforth refer to (\ref{uv}) as $CMA$
system.

The Lagrangian density for the original form (\ref{cma}) of the
complex Monge-Amp\`ere equation was suggested in \cite{yn}
\begin{equation}
L = \frac{1}{6} [u_1u_{\bar 1}u_{2\bar 2} + u_2u_{\bar 2}u_{1\bar
1}-u_1u_{\bar 2}u_{2\bar 1} - u_2u_{\bar 1}u_{1\bar 2}] +
\varepsilon u ,
 \label{cmalagrange}
\end{equation}
which in our new notation for independent variables becomes
\begin{eqnarray}
 & & L = \frac{1}{6}[(u_t^2+u_x^2)u_{w\bar w} + u_wu_{\bar
w}(u_{tt}+u_{xx})
 \label{Lxtw}
\\ & & \phantom{L =} \mbox{} - u_{\bar w}(u_t-iu_x)(u_{tw}+iu_{xw})
- u_{w}(u_t+iu_x)(u_{t\bar w}-iu_{x\bar w})] + \varepsilon u .
\nonumber
\end{eqnarray}
For our purposes, we choose the Lagrangian density for the
first-order $CMA$ system (\ref{uv}) to be linear in the time
derivatives of the unknowns $u_t$ and $v_t$:
\begin{eqnarray}
 L = \frac{1}{6}\{ (4vu_t-3v^2+u_x^2)u_{w\bar w} + u_wu_{\bar
w}(v_t+u_{xx}) - u_x(u_{\bar w}u_{xw}+u_wu_{x\bar w}) \nonumber
\\ \mbox{} - u_t\Bigl(u_{\bar w}(v_w+2iu_{xw}) + u_w(v_{\bar w}-2iu_{x\bar
w})\Bigr)\} + \varepsilon u
 \label{Luv}
\end{eqnarray}
which, after substituting $v=u_t$, coincides with our original
Lagrangian (\ref{Lxtw}) up to a total divergence.

\section{Symplectic two-form and first Hamiltonian structure}
\setcounter{equation}{0}
 \label{sec-Hamilton}

Since the Lagrangian density (\ref{Luv}) is linear in $u_t$ and
$v_t$, the canonical momenta
\begin{eqnarray}
 & & \pi_u = \frac{\partial L}{\partial u_t} = \frac{1}{6}
[4vu_{w\bar w} - u_{\bar w}(v_w+2iu_{xw}) - u_w(v_{\bar w}
-2iu_{x\bar w})] \nonumber
\\ & & \pi_v = \frac{\partial L}{\partial v_t} =
\frac{1}{6} u_wu_{\bar w}
 \label{pi_v}
\end{eqnarray}
cannot be inverted for the velocities $u_t$ and $v_t$ and the
Lagrangian is thus degenerate. Therefore, following Dirac
\cite{dirac}, we impose them as constraints
\begin{eqnarray}
\phi_u & = & \pi_u + \frac{1}{6} [-4vu_{w\bar w} + u_{\bar
w}(v_w+2iu_{xw}) + u_w(v_{\bar w} -2iu_{x\bar w})] = 0 \nonumber
 \\ \phi_v & = & \pi_v - \frac{1}{6} u_wu_{\bar w} = 0
 \label{constraints}
\end{eqnarray}
and calculate the Poisson brackets of the constraints (more
details of the procedure were given in \cite{nns})
\begin{equation}
K_{ik} = \left[ \phi_i(x,w,\bar w) , \phi_k(x',w',\bar w') \right]
\label{kik}
\end{equation}
organizing them in a matrix form. This yields us the inverse of
the Hamiltonian operator
\begin{equation}
 K =    \left(            \begin{array}{cr}
 (v_{\bar w}-iu_{x\bar w}) D_w + (v_w+iu_{xw}) D_{\bar w} + v_{w\bar w} & - u_{w\bar w} \\
  u_{w\bar w} & 0
\end{array}   \right)
\label{kmu}
\end{equation}
as an explicitly skew-symmetric local operator. A symplectic
2-form is a volume integral $\Omega = \int\limits_{V}\omega dx dw
d\bar w$ of the density
\begin{equation}
 \omega = \frac{1}{2} \, d u^i \wedge K_{ij} \, d u^j =
\frac{1}{2} (v_{\bar w}-iu_{x\bar w}) d u \wedge d u_w +
\frac{1}{2} (v_{w}+iu_{xw}) d u \wedge d u_{\bar w} + u_{w\bar w}
dv \wedge du
 \label{defomega}
\end{equation}
where $u^1 = u$ and $u^2 = v$. In $\omega$, under the sign of the
volume integral, we can neglect all the terms that are either
total derivatives or total divergencies due to suitable boundary
conditions on the boundary surface of the volume.

For the exterior differential of this 2-form we obtain
\begin{eqnarray}
& &  d \omega = -i d u_x \wedge d u_w \wedge d u_{\bar w} =
-(i/3)\Bigl( D_x(du \wedge du_w \wedge du_{\bar w})
 \label{domega}
\\ & & \mbox{} + D_w(du_x
\wedge du \wedge du_{\bar w}) + D_{\bar w}(du_x \wedge du_w \wedge
du)\Bigr) \iff 0 \nonumber
\end{eqnarray}
that is, a total divergence which is equivalent to zero, so that
the 2-form $\Omega$ is closed and hence symplectic. As a
consequence, the Jacobi's identity is satisfied for the
corresponding Hamiltonian operator $J_0 = K^{-1}$ obtained by
inverting $K$ in (\ref{kmu})
\begin{eqnarray}
\hspace{-15.8pt} & & J_0 =
 \label{J0}
  \\ \hspace{-15.8pt} & & \left(
 \begin{array}{cc}
0 & \frac{\textstyle 1}{\textstyle u_{w\bar w}}
\\[2mm] -\frac{\textstyle 1}{\textstyle u_{w\bar w}} & \frac{\textstyle v_{\bar w}-iu_{x\bar w}}{\textstyle 2u_{w\bar
w}^2} D_w + D_w \frac{\textstyle v_{\bar w}-iu_{x\bar
w}}{\textstyle 2u_{w\bar w}^2} + \frac{\textstyle
v_w+iu_{xw}}{\textstyle 2u_{w\bar w}^2} D_{\bar w} + D_{\bar w}
\frac{\textstyle v_w+iu_{xw}}{\textstyle 2u_{w\bar w}^2}
\end{array}
\right) \nonumber
\end{eqnarray}
that is explicitly skew-symmetric. The direct proof of the
Jacobi's identity for $J_0$ was also performed with the use of the
functional multi-vectors criterion of Olver \cite{olv}.

The Hamiltonian density was calculated as
\[ H_1 = \pi_u u_t + \pi_v v_t - L \]
with the result
\begin{equation}
 H_1 = \frac{1}{6}\Bigr[(3v^2-u_x^2)u_{w\bar w} - u_wu_{\bar w}u_{xx} +
u_x(u_{\bar w}u_{xw}+u_wu_{x\bar w})\Bigl] - \varepsilon u .
 \label{H1}
\end{equation}
$CMA$ system can now be written in the Hamiltonian form
\begin{equation}
\left(
\begin{array}{c}
u_t \\ v_t
\end{array}
\right) =  J_0 \left(
\begin{array}{c}
\delta_u H_1 \\ \delta_v H_1
\end{array}
\right)
 \label{Hamilton}
\end{equation}
where $H_1$ is given in (\ref{H1}) and $\delta_u$ and $\delta_v$
are Euler-Lagrange operators \cite{olv} with respect to $u$ and
$v$ applied to the Hamiltonian density $H_1$ (they correspond to
variational derivatives of the Hamiltonian functional
$\int\limits_V H_1 dV$).

\section{Conclusion}

We have discovered  a symplectic and Hamiltonian structure of the
complex Monge-Amp\`ere equation set into a two-component
evolutionary form.  This is the first step to obtain a
multi-Hamiltonian structure of $CMA$. The next step is to
construct a recursion operator for symmetries that, acting on the
first Hamiltonian operator, will generate a second Hamiltonian
operator and a bi-Hamiltonian representation of the complex
Monge-Amp\`ere equation. This work is now in progress and close to
final.

\section*{Acknowledgements}

The research of MBS is partly supported by the research grant from
Bogazici University Scientific Research Fund, research project No.
07B301. \\ MBS thanks D. Yaz{\i}c{\i} for useful discussions.


\begin{thebibliography}{99}
\bibitem{nns}
 Neyzi F,  Nutku Y and Sheftel M B 2005 {\it J. Phys. A: Math.
Gen.} {\bf 38} 8473--8485 ({\it Prperint} nlin.SI/0505030)
\bibitem{nsy}
 Neyzi F, Sheftel M B and Yazici D 2007 {\it Physics of Atomic
Nuclei} {\bf 70} 584--592
\bibitem{pleb}
 Plebanski J F 1975 {\it J. Math. Phys.} {\bf 16} 2395--402
\bibitem{magri}
 Magri F 1978 {\it J. Math. Phys.} {\bf 19}
1156--1162 \\ Magri F 1980 {\it Nonlinear Evolution Equations and
Dynamical Systems (Lecture Notes in Physics {\rm vol 120})} ed. M
Boiti, F Pempinelli and G Soliani (New York: Springer) p 233
\bibitem{yn}
 Nutku Y 2000 {\it Phys. Lett. A} {\bf 268} 293 ({\it Prpeprint}
hep-th/0004164)
\bibitem{dirac}
 Dirac  P A M 1964 {\it Lectures on Quantum
Mechanics (Belfer Graduate School of Science Monographs} series 2)
(New York: Yeshiva University Press)
\bibitem{olv}
Olver P 1986 {\it Applications of Lie Groups to Differential
Equations} (New York: Springer-Verlag)
 \end{thebibliography}
\end{document}